\documentclass[article, 11pt, onecolumn, superscriptaddress]{revtex4-1}

\usepackage{amsmath}
\usepackage{graphicx, hyperref,color}
\usepackage{ulem}
\usepackage{mathtools, nccmath}
\usepackage{titlesec}

\newcommand{\MPIHalle}{Max-Planck Institute of Microstructure Physics, Weinberg 2, 06120 Halle (Saale), Germany}
\newcommand{\MPIDresden}{Max Planck Institute for Chemical Physics of Solids, 00187 Dresden, Germany}
\newcommand{\ucsb}{University of California at Santa Barbara, Santa Barbara, California, USA}
\newcommand{\colombia}{Universidad del Norte, Barranquilla, Colombia}
\newcommand{\mainz}{Johannes von Gutenberg University of Mainz, Mainz, Germany}
\newcommand{\praguea}{Academy of Sciences of the Czech Republic, Praha, Czech Republic}
\newcommand{\pragueb}{Charles University in Prague, Czech Republic}
\newcommand{\oxford}{Oxford Department of Physics, Oxford, England}
\newcommand{\colorado}{Colorado School of Mines, Goldon, Colorado, USA}
\newcommand{\jhu}{Johns Hopkins University, Baltimore, Maryland, USA}

\begin{document}

\title{Anomalous, anomalous Hall effect in the layered, Kagome, Dirac semimetal KV$_3$Sb$_5$}

\author{Shuo-Ying Yang*}
\affiliation{\MPIHalle}

\author{Yaojia Wang*}
\affiliation{\MPIHalle}

\author{Brenden R. Ortiz}
\affiliation{\ucsb}

\author{ Defa Liu}
\affiliation{\MPIHalle}

\author{Jacob Gayles}
\affiliation{\MPIDresden}

\author{Elena Derunova}
\affiliation{\MPIHalle}

\author{Rafael Gonzalez-Hernandez}
\affiliation{\colombia}

\author{Libor Smejkal}
\affiliation{\mainz}
\affiliation{\praguea}
\affiliation{\pragueb}

\author{Yulin Chen}
\affiliation{\oxford}

\author{Stuart S. P. Parkin}
\affiliation{\MPIHalle}

\author{Stephen D. Wilson}
\affiliation{\ucsb}

\author{Eric S. Toberer}
\affiliation{\colorado}

\author{Tyrel McQueen}
\affiliation{\jhu}

\author{Mazhar N. Ali}\email{maz@berkeley.edu}
\affiliation{\MPIHalle}

\date{\today}

\begin{abstract}

\indent{}The electronic anomalous Hall effect (AHE), where charge carriers acquire a velocity component orthogonal to an applied electric field, is one of the most fundamental and widely studied phenomena in physics. There are several different AHE mechanisms known, and material examples are highly sought after, however in the highly conductive (skew scattering) regime the focus has centered around ferromagnetic metals. Here we report the observation of a giant extrinsic AHE in KV$_3$Sb$_5$, an exfoliable, Dirac semimetal with a Kagome layer of Vanadium atoms. Although there has been no reports of magnetic ordering down to 0.25 K \cite{ortiz2019new}, the anomalous Hall conductivity (AHC) reaches $\approx$ 15,507 $\Omega^{-1}$cm$^{-1}$ with an anomalous Hall ratio (AHR) of $\approx$ 1.8$ \%$; an order of magnitude larger than Fe \cite{hou2015multivariable}. Defying expectations from skew scattering theory, KV$_3$Sb$_5$ shows an enhanced skew scattering effect that scales quadratically, not linearly, with the longitudinal conductivity ($\sigma_{xx}$), opening the possibility of reaching an anomalous Hall angle (AHA) of 90$^{\circ}$ in \textit{metals}; an effect thought reserved for quantum anomalous Hall \textit{insulators}. This observation raises fundamental questions about the AHE and opens a new frontier for AHE (and correspondingly SHE) exploration, stimulating investigation in a new direction of materials, including metallic geometrically frustrated magnets, spin-liquid candidates, and cluster magnets.

\end{abstract}

\maketitle

\indent{}Since its discovery more than a century ago, the AHE has been extensively studied both theoretically and experimentally \cite{nagaosa2010anomalous}. Historically, spontaneous AHEs have been explored in materials with internally broken time-reversal symmetry (TRS), due to ferro- or ferrimagnetic ordering \cite{karplus1954hall, smit1955spontaneous, smit1958spontaneous, pugh1953hall}. Recently, there has been a surge of interest in the exploration of non-spontaneous AHEs, which require the application of an external magnetic field to break the TRS. The resulting Hall response is not commensurate with the magnitude of the applied field, thus making it distinct from the ordinary Hall effect (OHE). Such a non-spontaneous AHE has been seen in non-magnetic ZrTe$_5$ \cite{liang2018anomalous} as well as in dilute magnetically doped Kondo systems \cite{manyala2004large}. Large AHE can arise from a variety of effects and a particularly interesting limit to explore is when the anomalous Hall angle (AHA) approaches 90$^{\circ}$; one characteristic of the intrinsic quantum anomalous Hall effect (QAHE) observed in TRS breaking topological insulators \cite{yu2010quantized, chang2013experimental, liu2016quantum}. In these insulators, the anomalous Hall conductivity ($\sigma_{AHE}$) becomes modulated by the conductance quantum while the longitudinal conductivity ($\sigma_{xx}$) approaches zero, resulting in the AHA (tan$^{-1}$($\frac{\sigma_{AHE}}{\sigma_{xx}}$)) approaching 90$^{\circ}$ \cite{chang2015high, bestwick2015precise}.  

\indent{}The AHE can be broadly divided into two categories: intrinsic and extrinsic \cite{nagaosa2010anomalous}. The intrinsic AHE is governed by the electronic structure of a material that causes an electron to acquire a transverse momentum as it travels in-between scattering events \cite{karplus1954hall, sundaram1999wave, onoda2002topological, jungwirth2002anomalous}. The extrinsic AHEs, on the other hand, are dependent on electrons scattering off of sudden changes in the periodic potential of a crystal, caused by structural defects or chemical and magnetic impurities \cite{smit1955spontaneous, smit1958spontaneous, berger1970side}. Extrinsic AHEs can be further categorized into the ``dirty regime" (low conductivity, small scattering time, $\tau$) \cite{berger1964influence, berger1970side}, and the ``clean regime" (high conductivity, large scattering time) \cite{smit1955spontaneous, smit1958spontaneous} which is dominated by skew scattering and is the focus of this work. 

\indent{}Much effort has been dedicated to understanding the different mechanisms that can give rise to skew scattering AHEs, as illustrated in Fig. 1a. In clean ferromagnets with spontaneously ordered magnetic moments, such as Fe, an AHE can be induced by electrons deflecting transversely by nonmagnetic impurities \cite{dheer1967galvanomagnetic, tian2009proper, miyasato2007crossover}. In paramagnetic systems, such as ZnO/MnZnO, spin-dependent electron scattering on localized magnetic moments can give rise to an AHE \cite{maryenko2017observation}. Recently, scattering off of spin-clusters (local groups of coupled spins), has been proposed: tiled magnetic clusters, like magnetic atoms in a Kagome net, can generate an enhanced skew scattering potential \cite{ishizuka2019theory} and thus a large AHE. Triangular materials, like those often explored as geometrically frustrated magnets and spin liquid candidates, as well as other types of cluster magnets, are particularly likely to exhibit this type of AHE \cite{tatara2002chirality, kawamura2003anomalous, ishizuka2018spin, ishizuka2019theory}. However, the necessity of a combination of high conductivity in a spin-cluster lattice has contributed to its lack of experimental observation.

\indent{}In this work, we present the observation of a giant extrinsic AHE in KV$_3$Sb$_5$, which is a highly conductive Dirac semimetal with geometric frustration due to its vanadium Kagome net. The highly dispersive Dirac bands of KV$_3$Sb$_5$ are observed using angle resolved photoelectron spectroscopy (ARPES) as shown in Fig. 1b, along with the density functional theory (DFT) calculated bandstructure. The DFT and ARPES match well both in terms of band dispersion and the Fermi surface geometry with the experimentally determined Fermi level appearing to be slightly below the predicted level ($\approx$ -10 meV). Even though there has been no reports of magnetic ordering of KV$_3$Sb$_5$ down to 0.25 K \cite{ortiz2019new}, the AHC, at 2 K, reaches as high as $\approx $ 15,507 $\Omega^{-1}$cm$^{-1}$ with an AHR of $\approx$ 1.8\%; an order of magnitude larger than Fe \cite{hou2015multivariable} and one of the largest AHEs observed. Surprisingly, the observed AHE scales quadratically with $\sigma_{xx}$, deviating from the linear scaling predicted from current skew scattering theories. This is the first example of a giant AHE \textit{without} ferromagnetic ordering in a magnetic system and prompts investigations into previously unconsidered material families; particularly metallic geometrically frustrated magnets, spin liquid candidates, and cluster magnets. It also raises new questions on the fundamental theory regarding AHE mechanisms in the high conductivity regime and poses the possibility of realizing an AHA of 90$^{\circ}$ in metallic systems. This observation opens a new frontier for the AHE (and SHE) born from the intersection of geometrically frustrated/cluster magnets and topological semimetals/metals, inviting exploration not just from theoretical and experimental physicists, but also materials scientists and solid state chemists.

%PRESENTATION%%%%%%%%%%%%%%%%%%%%%%%%%%%%%%%%%%%%%%%%%%%%%%%%%%%%%%%%%%%%%%%%%%%%%%%%%%%%%%%%%%%%%%%%%%%%%%%%%%%%%%%%%%%%%%%%%%%%%%%%%%%%%%%%%%%%%

%RT - Kink 80K, MR, QO - low Mass carriers, HIGH CONDUCTIVITY, LOW-FIELD LINEAR MR, HIGH FIELD QUADRATIC%%%%%%%%%%%%%%

\indent{} KV$_3$Sb$_5$ crystallizes in the P6/mmm space group (SG: 191) and as shown in the inset of Fig. 1c, its stacking is comprised of a Kagome lattice of vanadium octahedrally coordinated by antimony with potassium intercalated between layers. Previous work by Ortiz \textit{et al.} found that the compound displays paramagnetic behavior at high temperatures, before undergoing a transition at 80 K to either a dilute trimerized state or a highly frustrated state with localized moments \cite{ortiz2019new}. Considering the vanadium Kagome net, geometrical frustration of the magnetic sublattice is expected. DFT + U calculations carried out by Ortiz \textit{et al.} comparing disordered AFM and ferrimagnetic ordering also support this expectation \cite{ortiz2019new}. Transport experiments on those same crystals were carried out here on a series of KV$_3$Sb$_5$ nanoflakes of different thicknesses. Fig. 1c shows the typical temperature dependence of $\rho$$_x$$_x$ for a 110 nm thick device (see supplementary information); with decreasing temperature, a kink is visible in $\rho$$_x$$_x$ around 80 K, corresponding to the known magnetization and heat capacity anomaly \cite{ortiz2019new}. At low temperature, the $\rho$$_x$$_x$ reaches $\approx$ 1.5 $\mu\Omega$ cm which is comparable to that of high purity bulk Bismuth \cite{behnia2007signatures}. The magnetoresistance (MR) at various temperatures is shown in Fig. 1d, with Shubnikov de Hass (SdH) oscillations clearly visible above 4 T. Below 3 T, the MR is linear while at higher field it adopts a standard quadratic dependance with $\mu$$_0$\textbf{H} (Fig. 1d inset, $\mu$$_0$\textbf{H} is the applied magnetic field). Fitting the quadratic field dependence, the average carrier mobility at 5 K is extracted to be $\approx$ 1,000 cm$^2$V$^-$$^1$s$^-$$^1$. The fast Fourier transforms (FFT) below 35 K reveals two identifiable periods at 34.6 T and 148.9 T, as shown in Fig. 1e. Tracking the temperature dependence of the FFT amplitude, the Lifshitz-Kosevich fit yields an effective mass of 0.125m$_e$ for carriers related to the 34.6 T orbit. Such a low effective mass corresponds well with the highly dispersive Dirac crossings near the Fermi level. Angle dependent MR and SdH oscillation analysis (see supplementary Fig. 2) shows the 34.6 T peak deviates from the 1/cos($\theta$) line below $\approx$ 20$^{\circ}$, implying the 34.6~T pocket is not strictly 2D. 

\indent{} The SdH oscillations are also clearly visible in the Hall response as shown in Fig. 2a, when the current is applied in the \textit{ab} plane and $\mu$$_0$\textbf{H} is applied along the \textit{c}-axis. In the high field region (above 10 T) and below 15 K there is a sudden change of habit resulting in a field independent Hall response; this may correspond to a spin-flop or other magnetic transition which changes symmetry and modifies the Fermi surface. Future investigations into the rich high-field magnetic and electronic properties of KV$_3$Sb$_5$ are necessary to elucidate the cause of this behavior. 

\indent{}In the low field region, highlighted by the blue shading in Fig. 2a, an anti-symmetric sideways ``S"-shape is observed, which is a characteristic of either an AHE or a two-band OHE. Below 35 K, $\rho_{xy}$ exhibits a second broad hump centered around 7 T, but as the temperature is increased, this hump is gradually lost and a one band linear field dependence is recovered (Fig. 2b inset). The ``S"-shaped Hall feature, however, persists throughout this change-over and remains visible at higher temperature where the Hall appears to be linear. This indicates that the high field behavior of the Hall is related to the two-band OHE and that the low-field ``S"-shape is related to an AHE. Within the one-band temperature range, the electron concentrations ($n_{e}$) and mobilities ($\mu_{e}$) are extracted from linear fitting of the OHE and shown in Fig. 2b (the simultaneous fitting of the two-band model with MR and Hall is not possible due to the linear MR behavior in this regime). As the temperature is lowered, $\mu_{e}$ monotonically increases while $n_{e}$ shows a minimum at around 65 K, which may be related to the magnetic transition mentioned above. Figure 2c shows the extracted $\rho^{AHE}_{xy}$ taken by subtracting the local linear OHE background. The magnitude of the AHE monotonically decreases with increasing temperature until it is lost at around 50 K. To precisely extract the AHC ($\sigma^{AHE}_{xy}$) when $\rho_{xy}$ $<$ $\rho_{xx}$ with no approximation, the Hall conductivity is first obtained by inverting the resistivity matrix, $\sigma_{xy}$ = $-\rho$$_x$$_y$/($\rho$$_x$$_x$$^2$+$\rho$$_x$$_y$$^2$). Afterwards, the local linear ordinary Hall conductivity background is subtracted, leaving the $\sigma^{AHE}_{xy}$ as shown in Fig. 2c inset.

\indent{} To further confirm the AHE nature of the low field anomaly, detailed angle-dependent measurements were carried out. Fig. 3a shows the $\sigma_{AHE}$ dependence on the angle of $\textbf{$\mu_{0}$H}$ relative to the applied electric field and the inset shows $\sigma_{AHE}$ against the cos($\theta$). The AHC is angle-independent until $\textbf{$\mu_{0}$H}$ is tilted away from the $z$-axis by about 30$^{\circ}$, after which it rapidly decreases until it reaches 0 at $\textbf{$\mu_{0}$H}$ $\parallel$  $\textbf{E}$. The fact that $\sigma_{AHE}$ does not linearly scale with the out-of-plane component of $\textbf{$\mu_{0}$H}$ solidifies its AHE origin. Furthermore, as expected from a real Hall response, the sign of the AHE flips when rotated past 90$^{\circ}$ (see supplementary Fig. 3). The extracted $\sigma_{AHE}$ for several devices with thicknesses ranging from 30 nm to 128 nm is plotted against each device's $\sigma^2_{xx}$ (which was varied by changing the temperature) in Fig. 3b. The skew scattering and intrinsic components of the AHE can be fitted to $\sigma_{AHE}$ = $\alpha\sigma^{-1}_{xx0}\sigma^{2}_{xx} + \textsl{b}$, where $\alpha$ is the skew constant, $\sigma_{xx0}$ is the residual resistivity, and \textsl{b} is the intrinsic AHC \cite{tian2009proper}. Samples 1, 2, and 3 were fabricated from freshly exfoliated crystals while samples 4, 5, and 6 were fabricated a few weeks after exfoliation. All devices follow a square dependence with $\alpha$ varying from 0.0075(2) to 0.0172(5); more than an order of magnitude larger than Fe and Ni (0.00149 \cite{hou2015multivariable, tian2009proper} and 0.0007 \cite{ye2012temperature}, respectively). The inset shows the extracted intrinsic $\sigma_{AHE}$ which average to positive 500 $\Omega^{-1}$cm$^{-1}$ for samples 1-3 and negative 325 $\Omega^{-1}$cm$^{-1}$ for samples 4-6. Both of these values, including the sign flip, are in agreement with the predicted intrinsic AHE values calculated via the Kubo formalism of integrating the Berry curvature over all occupied bands and assuming a net out-of-plane ferromagnetic order (see supplementary information), confirming the robustness of the AHE extraction. The AHR percentages ($\frac{\sigma_{AHE}}{\sigma_{xx}}\times100$) for various KV$_3$Sb$_5$ devices as well as for Fe \cite{miyasato2007crossover} are shown in Fig. 3c. Throughout the measured $\sigma_{xx}$ range, the AHR of KV$_3$Sb$_5$ rises monotonically with $\sigma_{xx}$, unlike Fe which has a decreasing AHR throughout its intrinsic region until $\sigma_{xx}$ $\approx0.6\times10^6$ $\Omega^{-1}$cm$^{-1}$, at which point its skew scattering mechanism begins to dominate. Due to its smaller skew constant, the rate of increase of its AHR is smaller compared to KV$_3$Sb$_5$. 
 
%Despite strong AFM coupling and a magnetic transition (as yet not fully understood) at $\approx$80 K, it shows no signs of magnetic order down to low temperature within the resolution limit of the experiment. It is believed that due to the triangular lattice, KV$_3$Sb$_5$ either adopts an AFM lattice which has not yet been resolved or is magnetically frustrated; in either case, it is not a Kondo system. Detailed study and analysis of the magnetic structure of KV$_3$Sb$_5$ will help understand the mechanism behind its giant AHE. 

%COMPARISON AND PROPOSAL OF NEW PHYSICS%%%%%%%%%%%%%%%%%%%%%%%%%%%%%%%%%%%%%%%%%%%%%%%%%%%%%%%%%%%%%%%%%%%%%%%%%%%%%%%%%%%%%%
\indent{} To compare the observed AHE of KV$_3$Sb$_5$ with that of other materials, $\sigma_{AHE}$ vs $\sigma_{xx}$ for a variety of materials spanning the various AHE regimes from the ``dirty regime" (localized hopping regime) through to the skew scattering regime \cite{miyasato2007crossover} are plotted in Fig. 4. The different scaling relationships for the localized hopping regime ($\sigma^{1.6}_{xx}$) and skew scattering regimes ($\sigma_{xx}$) as well as the quadratic scaling behavior of Fe ($\sigma^2_{xx}$) are shown by black dotted lines. Purple lines show three AHRs and their corresponding AHA values. The AHE in KV$_3$Sb$_5$, for devices of varying thickness with its scaling shown as the red dotted line, begins to dwarf the  AHE of most materials by $\sigma_{xx}$ $\approx$ 2 $\times$ 10$^5$ $\Omega^{-1}$cm$^{-1}$ due to its quadratic scaling. 

\indent{} With an AHE this large, there are few comparable systems to KV$_3$Sb$_5$ as most materials have AHEs on the order of 10$^2$ $\Omega^{-1}$cm$^{-1}$. It is interesting to understand what mechanism is giving rise to such giant skew scattering AHE. One possibility for the enhanced skew scattering effect in KV$_3$Sb$_5$ is the recently proposed ``spin cluster" mechanism by Nagoasa \textit{et al.} \cite{ishizuka2019theory}; a triangular spin cluster or tiled clusters as in a Kagome net can act like a ``compound magnetic scattering center" and generate an enhanced skew scattering potential when, due to an external field, a distortion of the local order results in a net magnetization \cite{ishizuka2019theory}. In fact, another recently discovered Dirac, Kagome system, Fe$_3$Sn$_2$ also has an AHE following quadratic scaling with a similar skew constant (0.013), but due to its low longitudinal conductivity, its magnitude remains an order smaller than in KV$_3$Sb$_5$. However, the ``spin cluster" theory predicts a $\sigma_{AHE}$ $\approx$ $\sigma_{xx}$ relationship, assuming the weak magneto-electron coupling limit and without including spin-orbit coupling (SOC) in the derivation. This is the same case as Fe, which is theoretically predicted to have AHC linearly increase with increasing $\sigma_{xx}$ but experimentally observed to have quadratic scaling. A theoretical treatment of Dirac crossings in spin cluster systems where the SOC is crucial, may yield the necessary scattering potential that recovers the quadratic dependence on $\sigma_{xx}$ \cite{crepieux2001theory, vzelezny2017spin}.

\indent{} A combination of enhanced skew scattering parameters and the quadratic scaling grant the KV$_3$Sb$_5$-like materials the potential to realize another fascinating effect: to achieve an AHA approaching 90$^{\circ}$ extrinsically, which has not been proposed or observed, to the best of our knowledge. Quadratic dependence of $\sigma_{AHE}$ means that the AHA increases quickly with increasing $\sigma_{xx}$, allowing a very large $\sigma_{AHE}$ at reasonable $\sigma_{xx}$. For example, extrapolating the evident quadratic scaling of KV$_3$Sb$_5$ shown in Fig. 4, an AHA = 45$^{\circ}$ is reached by $\sigma_{xx}$ of 5 x 10$^7$ $\Omega^{-1}$cm$^{-1}$. These are very large conductivities, but not implausible; the Dirac semimetal Cd$_3$As$_2$ \cite{liang2015ultrahigh}, Weyl semimetal NbAs \cite{zhang2019ultrahigh} and encapsulated graphene \cite{wang2013one} all are known to reach this conductivity regime. A similar extrapolation for Fe would require an unrealistic $\sigma_{xx}$ of $ > $10$^8$ $\Omega^{-1}$cm$^{-1}$. 
%Another Dirac, Kagome system, Fe$_3$Sn$_2$ was recently observed to also have an AHE following quadratic scaling with a similar skew constant (0.013), but due to its low longitudinal conductivity, its AHE remains an order of magnitude smaller than in KV$_3$Sb$_5$.      

Intuitively, however, one would expect a saturation of the AHA as one approaches large $\sigma_{xx}$ values, as the magnetic scattering events are expected to become increasingly rare when the system is exceptionally clean and defect free. However, in the case of materials with some form of \textit{backscattering} protection, $\sigma_{xx}$ can increase without requiring vanishing defects. This has been seen in the Weyl semimetal WTe$_2$, where the Fermi Surface has spin polarized pockets resulting in spin-flip protection which suppresses backscattering \cite{jiang2015signature}; and in the Dirac semimetal Cd$_3$As$_2$ where the quantum scattering lifetime is several orders of magnitude less than the backscattering lifetime, meaning that although the electrons are scattering often, they are not backscattering often \cite{liang2015ultrahigh}. In the limit of an 90$^{\circ}$ AHA due to skew scattering, the time between skew scattering events would need to be much less than the time between backscattering events, such that an electron traveling through a soup of magnetic scattering centers in an applied magnetic field would have a very high probability to undergo an orthogonal scattering event before undergoing a backscattering event. In the case of a high conductivity, enhanced skew scattering material like KV$_3$Sb$_5$, the addition of some form of backscattering suppression should allow extremely large AHAs to be realizable as one could envision the scenario of a high concentration of scattering centers that do not prohibit $\sigma_{xx}$ from increasing, but do provide a high probability of orthogonal scattering. Rigorous theoretical handling of this limit is needed.

%An increasing AHA implies that either a larger percentage of the injected electrons undergo an orthogonal scattering event, or that the Hall velocity of the electrons increases, but since the scattering events become increasingly rare as $\sigma_{xx}$ increases, one would expect the AHE to eventually deviate from the $\sigma^2_{xx}$ scaling.

%\indent{}The recipe for extremely large extrinsic AHAs, approaching an AHA of 90$^{\circ}$ (which, to the best of our knowledge, has not yet been proposed or observed in \textit{metals}), is to get both the quadratic $\sigma_{xx}$ scaling as well as the enhanced skew constant in a high conductivity system. In terms of materials, this means the combination of geometrically frustrated/spin-cluster magnetic sublattices i.e. triangular and Kagome sublattices, in topological semimetals/metals. Another Dirac, Kagome system, Fe$_3$Sn$_2$ was recently observed to also have an AHE following quadratic scaling with a similar skew constant (0.013), but due to its low longitudinal conductivity, its AHE remains an order of magnitude smaller than in KV$_3$Sb$_5$.      

%Also, low temperature magnetization versus field measurements reveal a weak saturation behavoir (w/o hysteresis), implying ferromagnetic coupling occurs.

%CONCLUSION%%%%%%%%%%%%%%%%%%%%%%%%%%%%%%%%%%%%%%%%%%%%%%%%%%%%%%%%%%%%%%%%%%%%%%%%%%%%%%%%%%%%%%%%%%%%%%%%%%%%%%%%%%%%%%%%%%%%%%%%%%%%

\indent{}In summary, a giant extrinsic AHE as large as 15,507 $\Omega^{-1}$ cm$^{-1}$ is observed in the exfoliable, Dirac semimetal KV$_3$Sb$_5$, which hosts a Kagome net of Vanadium atoms without magnetic ordering down to 0.25 K. Unexpectedly, the observed $\sigma_{AHE}$ is proportional to $\sigma^2_{xx}$ and, combined with an enhanced skew constant, the AHA rapidly rises with increasing $\sigma_{xx}$; a phenomena not seen before. This points to the possibility of extremely large AHAs at reasonable conductivity values including an AHA of 45$^{\circ}$ by $\approx$ 5 $\times$10$^7$ $\Omega^{-1}$ cm$^{-1}$, and an AHA $\approx$ 90 $^{\circ}$ if the skew scattering constant can be further enhanced by an order of magnitude. Materials with \textsl{S} $>$ 1/2 may result in even greater skew constants, further lowering the $\sigma_{xx}$ threshold for an AHA = 90$^{\circ}$. This raises fundamental questions about the extrinsic limits of the AHE as one of the effects of quantum anomalous Hall insulators may be replicated in a high conductive metal. We speculate that the Kagome sublattice in KV$_3$Sb$_5$ may be acting as tilted spin-clusters giving rise to enhanced skew scattering potentials according to a recent proposal by Nagaosa \textit{et al.}, and suggest future theoretical studies on understanding the coupling strength of Dirac-like electrons to the magnetic texture with SOC to reveal detailed scaling relations. Sister compounds RbV$_3$Sb$_5$ and CsV$_3$Sb$_5$ are expected to show a similar effect. Since these materials have weakly bound alkali earth interstitials, the Fermi level can be tuned through intercalation control; ionic liquid gating on few layer samples may be an ideal way to vary $\sigma_{xx}$ and explore the AHE response. Additionally, the high exfoliatability of these compounds makes them ideal platforms for thickness-dependent and monolayer exploration of the AHE and observing the cross-over from the extrinsic dominated AHE to the intrinsic dominated regime. Generally speaking, future work investigating more metallic geometrically frustrated and cluster magnet materials, particularly with topological band structures, may reveal many more enhanced skew scattering AHEs and help provide experimental basis for theoretical understanding of this phenomena. Lastly, the skew scattering \textit{spin} Hall effect arises from a similar mechanism as the skew scattering AHE, and therefore very large spin Hall angles may also be discovered in KV$_3$Sb$_5$ and other similar materials. This is another particularly important avenue of research as large spin Hall angles in highly conductive systems (and therefore low power) are extremely sought after for spintronic applications.

\textbf{Methods}

High quality single crystals of KV$_3$Sb$_5$ were synthesized from K (ingot, Alfa 99.8\%), V (powder, Sigma 99.9\%) and Sb (shot, Alfa 99.999\%) via the flux method as decribed by Ortiz et al \cite{ortiz2019new}. Flux mixtures containing 5 mol \% of KV$_3$Sb$_5$ were heated to 1000$^{\circ}$C, soaked for 24 hours, and then subsequently cooled at 2$^{\circ}$C per hour. KV$_3$Sb$_5$ crystals were structurally and chemically characterized by powder-XRD to confirm bulk purity, SEM-EDX for chemical analysis. \\

\indent{}A Quantum Design PPMS was used for transport measurements with Keithley 6221 and Keithley 2102 electronics. Hall measurements were taken in a 5-wire configuration while the magnetoresistance of KV$_3$Sb$_5$ samples was measured using the 4-point probe method. The rotator insert (Quantum Design) was used to tilt the angle between the magnetic field and the current. \\

\indent{}ARPES measurements were performed at Beamline I05 of the Diamond Light Source (DLS) using the Scienta R4000 analyzer. The angle and energy resolutions were $<$  0.2$^{\circ}$ and $<$ 15 meV, respectively. \\

\indent{}The electronic structure calculations were performed in the framework of density functional theory using the \textsc{wien2k} \cite{blaha1990full} code with a full-potential linearized augmented plane-wave and local orbitals [FP-LAPW + lo] basis \cite{blaha1990computer} together with the Perdew Burke Ernzerhof (PBE) parametrization of the generalized gradient approximation (GGA) as the exchange-correlation functional. The Fermi surface was plotted with the program \textit{Xcrysden}. \\

\textbf{Acknowledgements}
We acknowledge C. Cacho, T. Kim and P. Dudin for the support during the ARPES measurements. This research was supported by the Alexander von Humboldt Foundation Sofia Kovalevskaja Award, the MINERVA ARCHES Award, shared facilities of the UCSB MRSEC (NSF DMR 1720256). B.R.O. acknowledges support from the California NanoSystems Institute through the Elings Fellowship program. D.L. acknowledge the support from the Alexander von Humboldt Foundation. L.S. and R.G.-H. acknowledge the use of the supercomputer Mogon at JGU (hpc.uni-mainz.de), the Transregional Collaborative Research Center (SFB/TRR) 173 SPIN+X, the Grant Agency of the Czech Republic Grant No. 19-18623Y and support from the Institute of Physics of the Czech Academy of Sciences and the Max Planck Society through the Max Planck Partner Group programme.  \\

\textbf{Author Contributions}\
S.-Y.Y. and Y.W. were the lead researchers and contributed equally to this project. They measured the transport data and carried out the analysis. B.R.O. grew the samples and measured the magnetism. D.L. carried out the ARPES measurements. R.G-H. and L.S. performed and analysed the ab initio calculations. L.S., J.G. and E.D. carried out DFT analysis and provided theoretical support. Y.C., S.S.P.P., S.D.W., E.T., T.M. and M.N.A are the Principal Investigators. \\

\textbf{Competing Interests} 
The authors declare that they have no competing financial interests. \\

\textbf{Correspondence} Correspondence and requests for materials should be addressed to Mazhar N. Ali~(email: maz@berkeley.edu).

\bibliography{KVSb_citation}

\begin{thebibliography}{10}
\expandafter\ifx\csname url\endcsname\relax
  \def\url#1{\texttt{#1}}\fi
\expandafter\ifx\csname urlprefix\endcsname\relax\def\urlprefix{URL }\fi
\providecommand{\bibinfo}[2]{#2}
\providecommand{\eprint}[2][]{\url{#2}}

\bibitem{ortiz2019new}
\bibinfo{author}{Ortiz, B.~R.} \emph{et~al.}
\newblock \bibinfo{title}{New kagome prototype materials: discovery of kv3sb5,
  rbv3sb5, and csv3sb5}.
\newblock \emph{\bibinfo{journal}{Physical Review Materials}}
  \textbf{\bibinfo{volume}{3}}, \bibinfo{pages}{094407} (\bibinfo{year}{2019}).

\bibitem{hou2015multivariable}
\bibinfo{author}{Hou, D.} \emph{et~al.}
\newblock \bibinfo{title}{Multivariable scaling for the anomalous hall effect}.
\newblock \emph{\bibinfo{journal}{Physical review letters}}
  \textbf{\bibinfo{volume}{114}}, \bibinfo{pages}{217203}
  (\bibinfo{year}{2015}).

\bibitem{nagaosa2010anomalous}
\bibinfo{author}{Nagaosa, N.}, \bibinfo{author}{Sinova, J.},
  \bibinfo{author}{Onoda, S.}, \bibinfo{author}{MacDonald, A.~H.} \&
  \bibinfo{author}{Ong, N.~P.}
\newblock \bibinfo{title}{Anomalous hall effect}.
\newblock \emph{\bibinfo{journal}{Reviews of modern physics}}
  \textbf{\bibinfo{volume}{82}}, \bibinfo{pages}{1539} (\bibinfo{year}{2010}).

\bibitem{karplus1954hall}
\bibinfo{author}{Karplus, R.} \& \bibinfo{author}{Luttinger, J.}
\newblock \bibinfo{title}{Hall effect in ferromagnetics}.
\newblock \emph{\bibinfo{journal}{Physical Review}}
  \textbf{\bibinfo{volume}{95}}, \bibinfo{pages}{1154} (\bibinfo{year}{1954}).

\bibitem{smit1955spontaneous}
\bibinfo{author}{Smit, J.}
\newblock \bibinfo{title}{The spontaneous hall effect in ferromagnetics i}.
\newblock \emph{\bibinfo{journal}{Physica}} \textbf{\bibinfo{volume}{21}},
  \bibinfo{pages}{877--887} (\bibinfo{year}{1955}).

\bibitem{smit1958spontaneous}
\bibinfo{author}{Smit, J.}
\newblock \bibinfo{title}{The spontaneous hall effect in ferromagnetics ii}.
\newblock \emph{\bibinfo{journal}{Physica}} \textbf{\bibinfo{volume}{24}},
  \bibinfo{pages}{39--51} (\bibinfo{year}{1958}).

\bibitem{pugh1953hall}
\bibinfo{author}{Pugh, E.~M.} \& \bibinfo{author}{Rostoker, N.}
\newblock \bibinfo{title}{Hall effect in ferromagnetic materials}.
\newblock \emph{\bibinfo{journal}{Reviews of Modern Physics}}
  \textbf{\bibinfo{volume}{25}}, \bibinfo{pages}{151} (\bibinfo{year}{1953}).

\bibitem{liang2018anomalous}
\bibinfo{author}{Liang, T.} \emph{et~al.}
\newblock \bibinfo{title}{Anomalous hall effect in zrte 5}.
\newblock \emph{\bibinfo{journal}{Nature Physics}}
  \textbf{\bibinfo{volume}{14}}, \bibinfo{pages}{451} (\bibinfo{year}{2018}).

\bibitem{manyala2004large}
\bibinfo{author}{Manyala, N.} \emph{et~al.}
\newblock \bibinfo{title}{Large anomalous hall effect in a silicon-based
  magnetic semiconductor}.
\newblock \emph{\bibinfo{journal}{Nature materials}}
  \textbf{\bibinfo{volume}{3}}, \bibinfo{pages}{255} (\bibinfo{year}{2004}).

\bibitem{yu2010quantized}
\bibinfo{author}{Yu, R.} \emph{et~al.}
\newblock \bibinfo{title}{Quantized anomalous hall effect in magnetic
  topological insulators}.
\newblock \emph{\bibinfo{journal}{Science}} \textbf{\bibinfo{volume}{329}},
  \bibinfo{pages}{61--64} (\bibinfo{year}{2010}).

\bibitem{chang2013experimental}
\bibinfo{author}{Chang, C.-Z.} \emph{et~al.}
\newblock \bibinfo{title}{Experimental observation of the quantum anomalous
  hall effect in a magnetic topological insulator}.
\newblock \emph{\bibinfo{journal}{Science}} \textbf{\bibinfo{volume}{340}},
  \bibinfo{pages}{167--170} (\bibinfo{year}{2013}).

\bibitem{liu2016quantum}
\bibinfo{author}{Liu, C.-X.}, \bibinfo{author}{Zhang, S.-C.} \&
  \bibinfo{author}{Qi, X.-L.}
\newblock \bibinfo{title}{The quantum anomalous hall effect: Theory and
  experiment}.
\newblock \emph{\bibinfo{journal}{Annual Review of Condensed Matter Physics}}
  \textbf{\bibinfo{volume}{7}}, \bibinfo{pages}{301--321}
  (\bibinfo{year}{2016}).

\bibitem{chang2015high}
\bibinfo{author}{Chang, C.-Z.} \emph{et~al.}
\newblock \bibinfo{title}{High-precision realization of robust quantum
  anomalous hall state in a hard ferromagnetic topological insulator}.
\newblock \emph{\bibinfo{journal}{Nature materials}}
  \textbf{\bibinfo{volume}{14}}, \bibinfo{pages}{473} (\bibinfo{year}{2015}).

\bibitem{bestwick2015precise}
\bibinfo{author}{Bestwick, A.} \emph{et~al.}
\newblock \bibinfo{title}{Precise quantization of the anomalous hall effect
  near zero magnetic field}.
\newblock \emph{\bibinfo{journal}{Physical review letters}}
  \textbf{\bibinfo{volume}{114}}, \bibinfo{pages}{187201}
  (\bibinfo{year}{2015}).

\bibitem{sundaram1999wave}
\bibinfo{author}{Sundaram, G.} \& \bibinfo{author}{Niu, Q.}
\newblock \bibinfo{title}{Wave-packet dynamics in slowly perturbed crystals:
  Gradient corrections and berry-phase effects}.
\newblock \emph{\bibinfo{journal}{Physical Review B}}
  \textbf{\bibinfo{volume}{59}}, \bibinfo{pages}{14915} (\bibinfo{year}{1999}).

\bibitem{onoda2002topological}
\bibinfo{author}{Onoda, M.} \& \bibinfo{author}{Nagaosa, N.}
\newblock \bibinfo{title}{Topological nature of anomalous hall effect in
  ferromagnets}.
\newblock \emph{\bibinfo{journal}{Journal of the Physical Society of Japan}}
  \textbf{\bibinfo{volume}{71}}, \bibinfo{pages}{19--22}
  (\bibinfo{year}{2002}).

\bibitem{jungwirth2002anomalous}
\bibinfo{author}{Jungwirth, T.}, \bibinfo{author}{Niu, Q.} \&
  \bibinfo{author}{MacDonald, A.~H.}
\newblock \bibinfo{title}{Anomalous hall effect in ferromagnetic
  semiconductors}.
\newblock \emph{\bibinfo{journal}{Physical review letters}}
  \textbf{\bibinfo{volume}{88}}, \bibinfo{pages}{207208}
  (\bibinfo{year}{2002}).

\bibitem{berger1970side}
\bibinfo{author}{Berger, L.}
\newblock \bibinfo{title}{Side-jump mechanism for the hall effect of
  ferromagnets}.
\newblock \emph{\bibinfo{journal}{Physical Review B}}
  \textbf{\bibinfo{volume}{2}}, \bibinfo{pages}{4559} (\bibinfo{year}{1970}).

\bibitem{berger1964influence}
\bibinfo{author}{Berger, L.}
\newblock \bibinfo{title}{Influence of spin-orbit interaction on the transport
  processes in ferromagnetic nickel alloys, in the presence of a degeneracy of
  the 3d band}.
\newblock \emph{\bibinfo{journal}{Physica}} \textbf{\bibinfo{volume}{30}},
  \bibinfo{pages}{1141--1159} (\bibinfo{year}{1964}).

\bibitem{dheer1967galvanomagnetic}
\bibinfo{author}{Dheer, P.}
\newblock \bibinfo{title}{Galvanomagnetic effects in iron whiskers}.
\newblock \emph{\bibinfo{journal}{Physical Review}}
  \textbf{\bibinfo{volume}{156}}, \bibinfo{pages}{637} (\bibinfo{year}{1967}).

\bibitem{tian2009proper}
\bibinfo{author}{Tian, Y.}, \bibinfo{author}{Ye, L.} \& \bibinfo{author}{Jin,
  X.}
\newblock \bibinfo{title}{Proper scaling of the anomalous hall effect}.
\newblock \emph{\bibinfo{journal}{Physical review letters}}
  \textbf{\bibinfo{volume}{103}}, \bibinfo{pages}{087206}
  (\bibinfo{year}{2009}).

\bibitem{miyasato2007crossover}
\bibinfo{author}{Miyasato, T.} \emph{et~al.}
\newblock \bibinfo{title}{Crossover behavior of the anomalous hall effect and
  anomalous nernst effect in itinerant ferromagnets}.
\newblock \emph{\bibinfo{journal}{Physical review letters}}
  \textbf{\bibinfo{volume}{99}}, \bibinfo{pages}{086602}
  (\bibinfo{year}{2007}).

\bibitem{maryenko2017observation}
\bibinfo{author}{Maryenko, D.} \emph{et~al.}
\newblock \bibinfo{title}{Observation of anomalous hall effect in a
  non-magnetic two-dimensional electron system}.
\newblock \emph{\bibinfo{journal}{Nature communications}}
  \textbf{\bibinfo{volume}{8}}, \bibinfo{pages}{14777} (\bibinfo{year}{2017}).

\bibitem{ishizuka2019theory}
\bibinfo{author}{Ishizuka, H.} \& \bibinfo{author}{Nagaosa, N.}
\newblock \bibinfo{title}{Theory of giant skew scattering by spin cluster}.
\newblock \emph{\bibinfo{journal}{arXiv preprint arXiv:1906.06501}}
  (\bibinfo{year}{2019}).

\bibitem{tatara2002chirality}
\bibinfo{author}{Tatara, G.} \& \bibinfo{author}{Kawamura, H.}
\newblock \bibinfo{title}{Chirality-driven anomalous hall effect in weak
  coupling regime}.
\newblock \emph{\bibinfo{journal}{Journal of the Physical Society of Japan}}
  \textbf{\bibinfo{volume}{71}}, \bibinfo{pages}{2613--2616}
  (\bibinfo{year}{2002}).

\bibitem{kawamura2003anomalous}
\bibinfo{author}{Kawamura, H.}
\newblock \bibinfo{title}{Anomalous hall effect as a probe of the chiral order
  in spin glasses}.
\newblock \emph{\bibinfo{journal}{Physical review letters}}
  \textbf{\bibinfo{volume}{90}}, \bibinfo{pages}{047202}
  (\bibinfo{year}{2003}).

\bibitem{ishizuka2018spin}
\bibinfo{author}{Ishizuka, H.} \& \bibinfo{author}{Nagaosa, N.}
\newblock \bibinfo{title}{Spin chirality induced skew scattering and anomalous
  hall effect in chiral magnets}.
\newblock \emph{\bibinfo{journal}{Science advances}}
  \textbf{\bibinfo{volume}{4}}, \bibinfo{pages}{eaap9962}
  (\bibinfo{year}{2018}).

\bibitem{behnia2007signatures}
\bibinfo{author}{Behnia, K.}, \bibinfo{author}{Balicas, L.} \&
  \bibinfo{author}{Kopelevich, Y.}
\newblock \bibinfo{title}{Signatures of electron fractionalization in
  ultraquantum bismuth}.
\newblock \emph{\bibinfo{journal}{Science}} \textbf{\bibinfo{volume}{317}},
  \bibinfo{pages}{1729--1731} (\bibinfo{year}{2007}).

\bibitem{ye2012temperature}
\bibinfo{author}{Ye, L.}, \bibinfo{author}{Tian, Y.}, \bibinfo{author}{Jin, X.}
  \& \bibinfo{author}{Xiao, D.}
\newblock \bibinfo{title}{Temperature dependence of the intrinsic anomalous
  hall effect in nickel}.
\newblock \emph{\bibinfo{journal}{Physical Review B}}
  \textbf{\bibinfo{volume}{85}}, \bibinfo{pages}{220403}
  (\bibinfo{year}{2012}).

\bibitem{crepieux2001theory}
\bibinfo{author}{Cr{\'e}pieux, A.} \& \bibinfo{author}{Bruno, P.}
\newblock \bibinfo{title}{Theory of the anomalous hall effect from the kubo
  formula and the dirac equation}.
\newblock \emph{\bibinfo{journal}{Physical Review B}}
  \textbf{\bibinfo{volume}{64}}, \bibinfo{pages}{014416}
  (\bibinfo{year}{2001}).

\bibitem{vzelezny2017spin}
\bibinfo{author}{{\v{Z}}elezn{\`y}, J.}, \bibinfo{author}{Zhang, Y.},
  \bibinfo{author}{Felser, C.} \& \bibinfo{author}{Yan, B.}
\newblock \bibinfo{title}{Spin-polarized current in noncollinear
  antiferromagnets}.
\newblock \emph{\bibinfo{journal}{Physical review letters}}
  \textbf{\bibinfo{volume}{119}}, \bibinfo{pages}{187204}
  (\bibinfo{year}{2017}).

\bibitem{liang2015ultrahigh}
\bibinfo{author}{Liang, T.} \emph{et~al.}
\newblock \bibinfo{title}{Ultrahigh mobility and giant magnetoresistance in the
  dirac semimetal cd3as2}.
\newblock \emph{\bibinfo{journal}{Nature materials}}
  \textbf{\bibinfo{volume}{14}}, \bibinfo{pages}{280} (\bibinfo{year}{2015}).

\bibitem{zhang2019ultrahigh}
\bibinfo{author}{Zhang, C.} \emph{et~al.}
\newblock \bibinfo{title}{Ultrahigh conductivity in weyl semimetal nbas
  nanobelts}.
\newblock \emph{\bibinfo{journal}{Nature materials}}
  \textbf{\bibinfo{volume}{18}}, \bibinfo{pages}{482} (\bibinfo{year}{2019}).

\bibitem{wang2013one}
\bibinfo{author}{Wang, L.} \emph{et~al.}
\newblock \bibinfo{title}{One-dimensional electrical contact to a
  two-dimensional material}.
\newblock \emph{\bibinfo{journal}{Science}} \textbf{\bibinfo{volume}{342}},
  \bibinfo{pages}{614--617} (\bibinfo{year}{2013}).

\bibitem{jiang2015signature}
\bibinfo{author}{Jiang, J.} \emph{et~al.}
\newblock \bibinfo{title}{Signature of strong spin-orbital coupling in the
  large nonsaturating magnetoresistance material wte 2}.
\newblock \emph{\bibinfo{journal}{Physical review letters}}
  \textbf{\bibinfo{volume}{115}}, \bibinfo{pages}{166601}
  (\bibinfo{year}{2015}).

\bibitem{blaha1990full}
\bibinfo{author}{Blaha, P.}, \bibinfo{author}{Schwarz, K.},
  \bibinfo{author}{Sorantin, P.} \& \bibinfo{author}{Trickey, S.}
\newblock \bibinfo{title}{Full-potential, linearized augmented plane wave
  programs for crystalline systems}.
\newblock \emph{\bibinfo{journal}{Computer Physics Communications}}
  \textbf{\bibinfo{volume}{59}}, \bibinfo{pages}{399--415}
  (\bibinfo{year}{1990}).

\bibitem{blaha1990computer}
\bibinfo{author}{Blaha, P.}
\newblock \bibinfo{title}{Computer code wien2k (vienna university of
  technology, 2002), improved and updated unix version of the original p.
  blaha, k. schwarz, p. sorantin, sb rickey}.
\newblock \emph{\bibinfo{journal}{Comput. Phys. Commun}}
  \textbf{\bibinfo{volume}{59}}, \bibinfo{pages}{399} (\bibinfo{year}{1990}).

\end{thebibliography}

\begin{figure*}[t]
\includegraphics[width=0.9\textwidth]{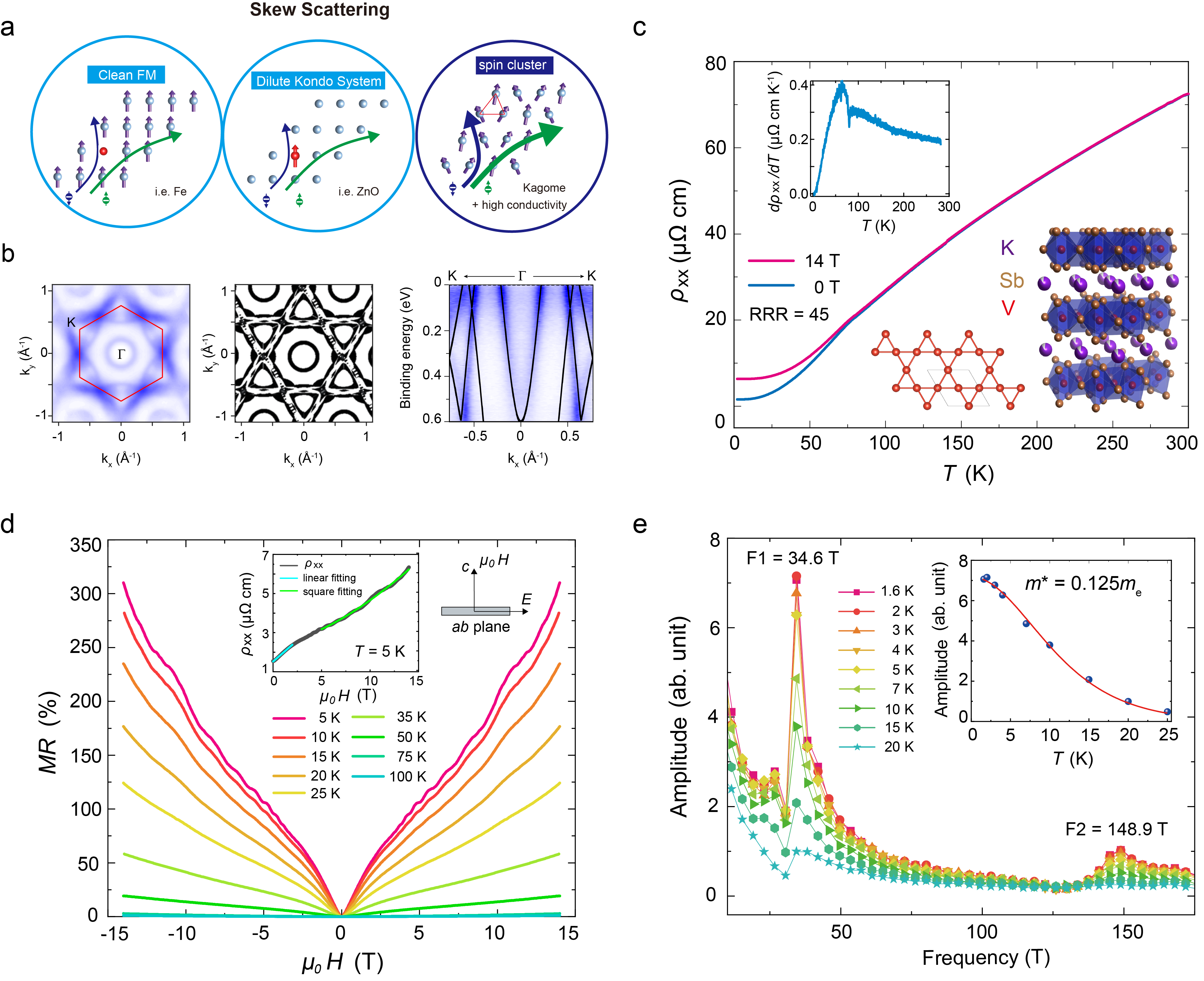}
\caption{Skew scattering mechanisms, basic ARPES and transport characteristics of KV$_3$Sb$_5$. (a) Schematic representation of three different skew scattering mechanisms, including clean ferromagnetic model, Kondo model and spin cluster model. (b) Left: experimentally measured Fermi surface of KV$_3$Sb$_5$. The hexagonal Brillouin zone is marked by the red line. Middle: DFT calculated Fermi surface. Right: band dispersion along K-$\Gamma$-K direction overlayed with the ARPES measurement. (c) Temperature dependence of longitudinal resistivity in zero field and in a field of 14 T. The inset on the upper left shows the temperature dependence of differential longitudinal resistivity in zero field, in which a kink at 80 K corresponds to the magnetic transition. The inset on the bottom right shows the crystal structure of KV$_3$Sb$_5$ and its Kagome lattice. (d) Magnetoresistance measured at various temperatures, exhibiting linear field dependence below 3 T and quadratic field dependence at higher field (inset). (e) Extracted FFT frequency showing two identifiable periods of 34.6 T and 148.9 T. The inset shows the Lifshitz-Kosevich fit of the 34.6 T orbit with an extracted effective mass of 0.125m$_e$.} \label{Figure_1}
\end{figure*}
\newpage

%FIGURE 2%%%%%%%%%%%%%%%%%%%%%%%%%%%%%%%%%%%%%%%%%%%%%%%%%%%%%%%%%%%%%%%%%%%%%%%%%%%%%%%%%%%%%%%%%%%%%%%%%%%%%%%%%%%%%%%%%%%%%%%%%%%%%%%%%%%%%%%%%%%%%%

\begin{figure}[h]
	\includegraphics[width=1\textwidth]{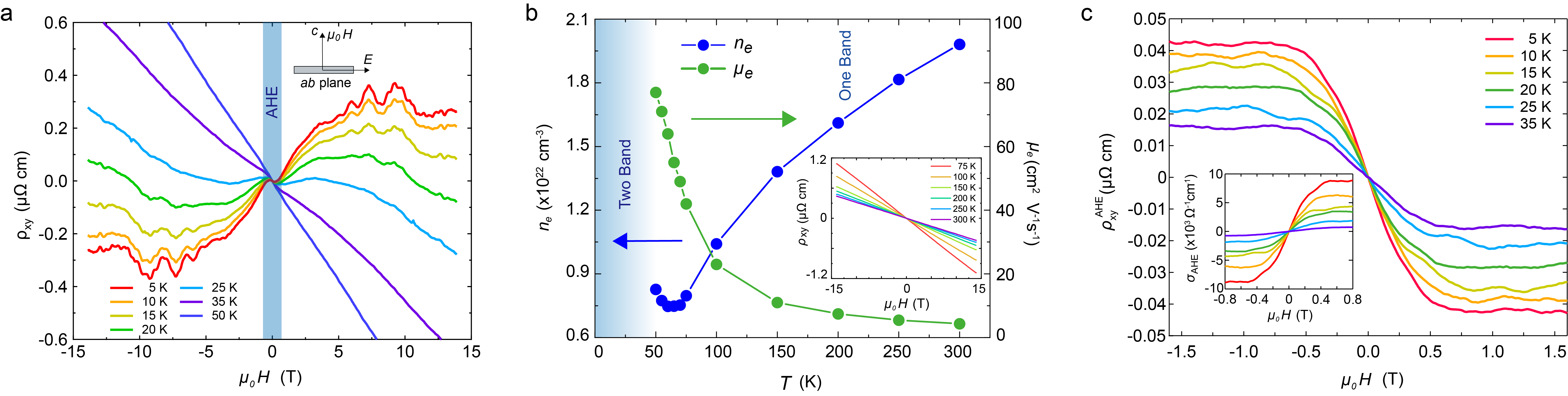}
	\caption{Hall effects in KV$_3$Sb$_5$. (a) The Hall resistivity of KV$_3$Sb$_5$ with the current applied in the \textit{ab} plane and the magnetic field applied along the $c$-axis. The AHE shows up as anti-symmetric ``S''-shape in the low field region for all temperature below 50 K. At low temperatures and high field regime, the Hall resistivity exhibit a typical two-band behavior. (b) Extracted electron carrier concentration and mobility in the one-band regime. Inset: the Hall response of KV$_3$Sb$_5$ above 75 K. (c) Extracted $\rho^{AHE}_{xy}$ taken by subtracting the local linear ordinary Hall background at various temperatures. The inset shows the converted $\sigma^{AHE}_{xy}$ at various temperatures by inverting the resistivity tensor.}
	\label{Figure_2}
\end{figure}
\newpage

%FIGURE 3%%%%%%%%%%%%%%%%%%%%%%%%%%%%%%%%%%%%%%%%%%%%%%%%%%%%%%%%%%%%%%%%%%%%%%%%%%%%%%%%%%%%%%%%%%%%%%%%%%%%%%%%%%%%%%%%%%%%%%%%%%%%%%%%%%%%%%%%%%%%%%%%%

\newpage
\begin{figure}[h]
	\includegraphics[width=1\textwidth]{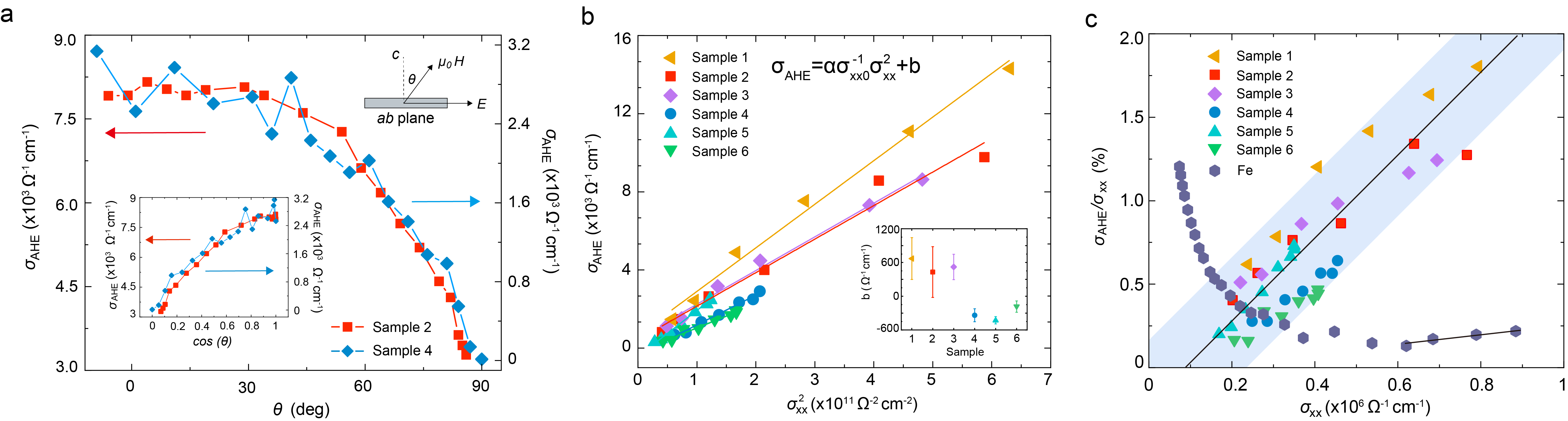}
	\caption{Anomalous Hall effect in KV$_3$Sb$_5$. (a) Angular dependence of $\sigma_{AHE}$ as the $\textbf{$\mu_{0}$H}$ is tilted from out-of-plane to in-plane. The inset shows the $\sigma_{AHE}$ against cos($\theta$). (b) Extracted $\sigma_{AHE}$ versus $\sigma^2_{xx}$ for various devices with thickness ranging from 30 nm to 128 nm. Solid lines are fittings to the equation shown in the inset to extract the skew scattering constant ($\alpha$) and intrinsic AHC ($b$) for each device. The inset shows the extracted intrinsic $\sigma_{AHE}$ for all six devices. Larger error is seen for samples 1-3 due to the size of the dominating extrinsic component. (c) The ratio between $\sigma_{AHE}$ and $\sigma_{xx}$ for six KV$_3$Sb$_5$ devices as well as for Fe. The black lines guide the eye to illustrate the increasing tendency of $\sigma_{AHE}/ \sigma_{xx}$ for KV$_3$Sb$_5$ and for Fe.}
	\label{Figure_3}
\end{figure}
\newpage

%FIGURE 4%%%%%%%%%%%%%%%%%%%%%%%%%%%%%%%%%%%%%%%%%%%%%%%%%%%%%%%%%%%%%%%%%%%%%%%%%%%%%%%%%%%%%%%%%%%%%%%%%%%%%%%%%%%%%%%%%%%%%%%%%%%%%%%%%%%%%%%%%%%%%%%%%%%

\begin{figure}[h]
	\includegraphics[width=0.8\textwidth]{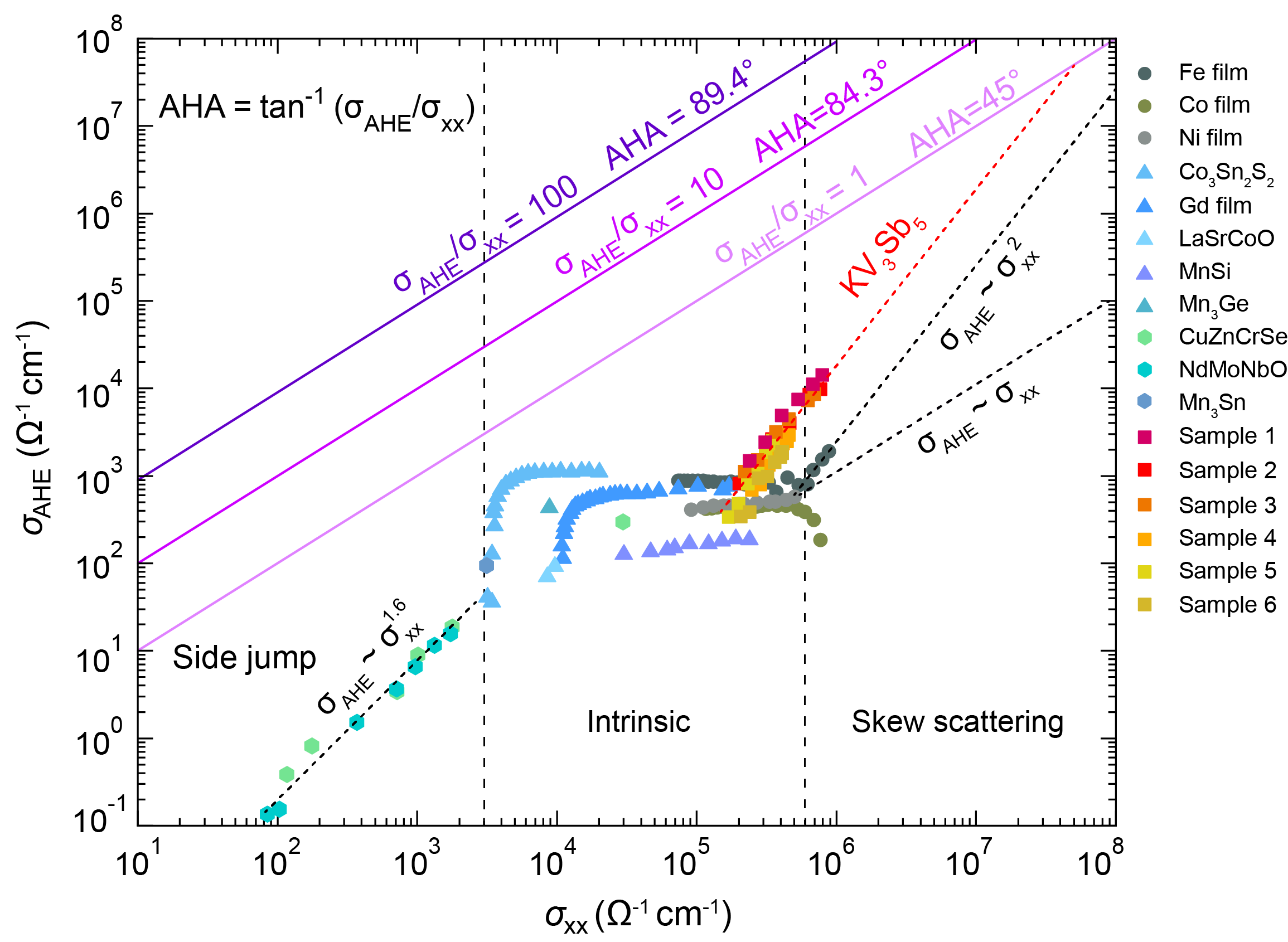}
	\caption{Map of anomalous Hall effects for various materials. $\sigma_{AHE}$ vs $\sigma_{xx}$ for a variety of materials spanning the various AHE regimes from the side-jump (SJ) mechanism through the intrinsic and skew scattering regimes.}
	\label{Figure_4}
\end{figure} 
\newpage

\end{document}